\documentstyle[useAMS,graphics,psfig]{mn2e}

\def\zabs{$z_{\rm abs}$}

\def\lya{Ly$\alpha$ }

\def\h2{H$_2$}

\def\ovi{O~{\sc vi}~ }

\def\nv{N~{\sc v}~ }

\def\cii{C~{\sc ii}~}

\def\civ{C~{\sc iv}~}
\def\civa{C~{\sc iv}$\lambda$1548~ }
\def\civb{C~{\sc iv}$\lambda$1550~ }

\def\siii{Si~{\sc ii}~}

\def\siiv{Si~{\sc iv}~}

\def\kms{km~s$^{-1}$}
\begin{document}
%
\title[Collimated flow driven by radiative pressure]{A collimated 
flow driven by radiative pressure from the nucleus of quasar Q~1511+091\thanks{Based 
on observations carried out at the European Southern Observatory (ESO) under
prog. ID No. 65.P-0038 and 65.O-0063 with the UVES spectrograph on the 
Very Large Telescope (VLT) at Cerro Paranal Observatory in Chile.}
}
%
%
\author[ R. Srianand, Patrick Petitjean, Cedric Ledoux \& Cyril Hazard]{
R. Srianand$^{1}$, Patrick Petitjean$^{2,3}$, Cedric Ledoux$^{4}$
\& Cyril Hazard$^{5}$\\ 
   ${}^1$ IUCAA, Post Bag 4, Ganeshkhind, Pune 411 007, India -
   email: anand@iucaa.ernet.in\\
   ${}^2$ Institut d'Astrophysique de Paris -- CNRS, 98bis Boulevard 
   Arago, F-75014 Paris, France - email: petitjean@iap.fr\\
   ${}^3$ LERMA, Observatoire de Paris, 61 avenue de l'Observatoire, F-75014 - 
Paris, France\\
   ${}^4$ European Southern Observatory, Alonso de C\'ordova 3107, Casilla 19001, 
Vitacura, Santiago, Chile- email: cledoux@eso.org\\
   ${}^5$ Institute of Astronomy, Madingley Road, CB3OHA Cambridge, United Kingdom\\
}
%
\date{Typeset \today ; Received / Accepted}
\maketitle
%
\begin{abstract}
High velocity outflows from quasars are revealed by the absorption 
signatures they produce in the spectrum of the quasar. 
Clues on the nature and origin of these flows are important for our 
understanding of the dynamics of gas in the central regions of the 
Active Galactic Nucleus (AGNs) but also of the metal enrichment of the 
intergalactic space. Line radiation  pressure has often been suggested 
to be an important process in driving these outflows, however no 
convincing evidence has been given so far. 
Here we report observation of a highly structured flow, 
toward Q~1511+091, where the velocity separations between distinct components 
are similar to O~{\sc vi}, N~{\sc v} and C~{\sc iv} doublet splittings with 
some of the profiles matching perfectly. This strongly favors the idea that 
the absorbing clumps originate at similar physical location and are driven 
by radiative acceleration due to resonance lines. The complex 
absorption can be understood if the flow is highly collimated so that the 
different optically thick clouds are aligned and cover the same region of 
the background source. One component shows saturated H~{\sc i} Lyman series
lines together with absorptions from excited levels from C~{\sc ii} and 
Si~{\sc ii} but covers only 40\% of the source of continuum. 
The fact that clouds cover only part of the small continuum source
implies that the flow is located very close to it.
\end{abstract}
\begin{keywords}
{\em Quasars:} absorption lines -- 
{\em Quasars:} individual: Q\,1511$+$091
\end{keywords}
\section{Introduction}
\begin{figure*}
\centerline{\vbox{
\psfig{figure=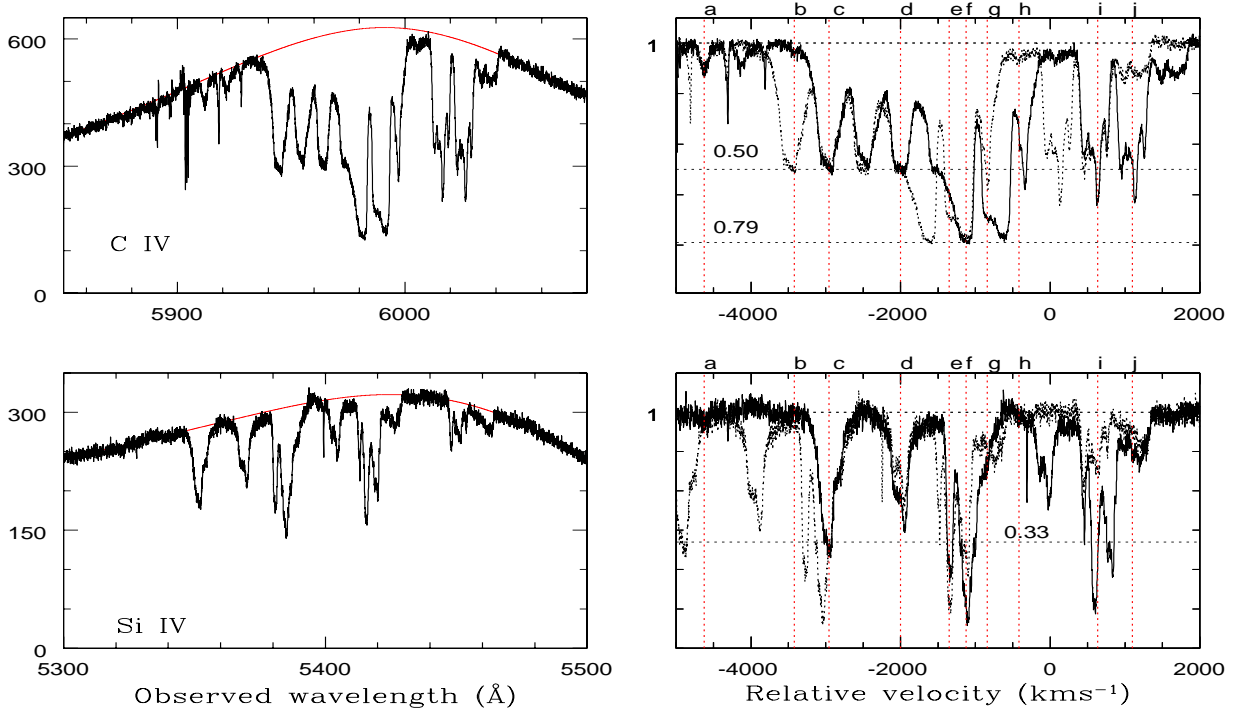,height=10.cm,width=18.cm,angle=270}
}}
\caption[]{
The left-hand side panels show the fit to the  effective
continuum (continuum plus broad emission line) of QSO 
spectrum in the \civ (top) and \siiv (bottom) emission line
regions. 
Right-hand side panels show the corresponding normalised fluxes on a 
velocity scale with respect to $z$~=~2.878. The y-axis in the left
and right panels give the observed spectrum (without flux calibration) and 
normalised flux respectively.
The continuous and dotted 
profiles are for the first (C~{\sc iv}$\lambda1548$ and Si~{\sc iv}$\lambda$1393) and second
(C~{\sc iv}$\lambda1550$ and Si~{\sc iv}$\lambda$1402) members of the 
doublets respectively. The horizontal dashed lines in each panels give the residual flux 
expected for different covering factors
(as indicated above each line). If the absorbing gas covers the continuum 
source only then the expected total covering factor of the
background source is $\sim 0.5$ and 
$\sim0.70$ 
for the \civ and \siiv lines respectively. These values
are obtained using the relative flux ratios of emission lines 
and continuum at the position of the absorption line from low
dispersion spectra of Sargent 
et al (1988).
}
\label{fig1}
\end{figure*}

Comprehensive understanding of the nature and origin of gas flows in
the very central regions of AGNs is elusive despite decades of 
intensive investigation. Presence of outflows with very large 
velocities ($\sim$20000~km~s$^{-1}$) is revealed by the broad 
as well as narrow absorption lines (BALs and NALs) observed
in the spectrum of AGNs. 
It has been suggested for long that the momentum transfered through 
resonance absorption lines could play an important
role in driving these flows as it does in the case of
O star winds (Castor, Abbott \& Klein 1975; Arav, 
Li, \& Begelman 1994). However, signatures of acceleration, though  
claimed to be noted in nearby Seyfert galaxy NGC~3515 
(Hutchings et al. 2001), are not generic. 
\par\noindent
A possible way to reveal line-driven acceleration 
is the presence of absorption features with velocity separation similar 
to doublet and/or multiplet splittings (a consequence of the so-called 
``line-locking'' effect). Such a structure in absorption is achieved when 
there is a reduction in the flux that drives the far away gas element due 
to line absorption produced by the gas closer to the QSO (Scargle 1973; 
Braun \& Milgrom 1989). 
For higher efficiency, the gas from different elements 
of the flow must cover the same region of the background source. Since 
the gas producing intrinsic absorption is known to cover the background source 
only partially (Petitjean et al. 1994; Hamann 1997; Barlow \& Sargent 1997),
line-locking can be achieve in this case if the 
flow is somehow collimated. This may be one of the reasons why tentative 
evidences for line-locking (Foltz et al. 1987; Srianand 2000;
Srianand \& Petitjean 2001; Vilkoviskiji \& Irwin 2001) as well as double 
trough (the so-called Lyman-alpha ghost) seen in the mean profile of 
BALs (Weymann et al. 1991; Korista et al. 1993; Arav, Li \&
Begelman 1994) are not completely convincing.
In this paper we present clear evidence for a flow
that is highly structured
through radiative pressure modulated by absorption in different 
resonance lines. 
\section{Observations}
\begin{table}
\caption {Covering factor of the absorbing gas}
\begin{tabular}{cclc}
\hline
\multicolumn{1}{c}{Component}&\multicolumn{1}{c}{\zabs}&\multicolumn{1}{c}{species}&\multicolumn{1}{c}{${\rm f_c}$}\\
\hline
$a$      & 2.8186  & C~{\sc iv}  &0.10$\pm$0.04\\
$c$      & 2.8398  & C~{\sc iv}  &0.48$\pm$0.02\\
$d$      & 2.8523  & Ly~$\alpha^b$ &0.30$\pm$0.01\\
         &         & Ly~series   &0.30$-$0.40  \\
         &         & C~{\sc iv}  &0.50$\pm$0.01\\
         &         & Si~{\sc iv} &0.30$\pm$0.05\\
         &         & N~{\sc v}$^b$&0.46$\pm$0.01\\
$e$      & 2.8607  & Ly~$\alpha^a$ &$\le 0.42$  \\
         &         & Ly~series   &0.40$-$0.45  \\
         &         & Al~{\sc iii}&0.43$\pm$0.03\\
         &         & C~{\sc iv}  &$\le 0.57$   \\
         &         & Si~{\sc iv} &$\le 0.49$   \\
$f$      & 2.8641  & C~{\sc iv}  &0.77$\pm$0.01\\
         &         & N~{\sc v}  &0.75$\pm$0.02\\
$j$(blue)& 2.8908  & Ly~$\alpha$ &$\ge0.15$    \\
	 &         & C~{\sc iv}  &$\ge0.10$    \\
	 &         & Si~{\sc iv} &0.28$\pm$0.12\\
         &         & N~{\sc v}   &0.22$\pm$0.03\\
$j$(red) & 2.8934  & Ly~$\alpha$ &$\ge0.15$    \\
         &         & C~{\sc iv}  &$\ge0.10$    \\
         &         & Si~{\sc iv} &0.17$\pm$0.07\\
         &         & N~{\sc v}   &0.18$\pm$0.02\\
\hline
\multicolumn{4}{l}{${^b}$ Line saturation is assumed}
\end{tabular}
\label{tab1}
\end{table}
We have used the Ultra-violet and Visible Echelle Spectrograph 
UVES (D'Odorico et al. 2000) mounted on the ESO KUEYEN 8.2~m telescope at 
the Paranal observatory on April 4-7 2000 to obtain a high-spectral 
resolution spectrum of the $z_{\rm em}$~=~2.878 and $m_{\rm V}$~=~17 
quasar Q~1511+091. Two standard dichroic settings have been used to 
observe with both arms of the spectrograph at the same time. The final 
wavelength range is 3260$-$10,000~\AA. The slit width was 1~arcsec resulting 
in a resolution of $\sim$45000. The total exposure time was 10 hours in 
seeing conditions better than 0.8~arcsec full width at half maximum. 
The data were reduced using the UVES pipeline, a set of procedures 
implemented in a dedicated context of MIDAS, the ESO data reduction package. 
The main characteristics of the pipeline is to perform a precise inter-order 
background subtraction for science frames and master flat-fields, and to 
allow for an optimal extraction of the object signal rejecting cosmic 
ray impacts and performing sky-subtraction at the same time. The reduction 
is checked step by step. Wavelengths were corrected to vacuum-heliocentric 
values and individual 1D spectra were combined together. This resulted 
in a S/N ratio per pixel of 25 at $\sim$3600~\AA~ and 80 at $\sim$6000~\AA. 
Typical errors in the wavelength calibration are less than 
$\sim$0.5~km~s$^{-1}$. 
\begin{figure*}
\centerline{\vbox{
\psfig{figure=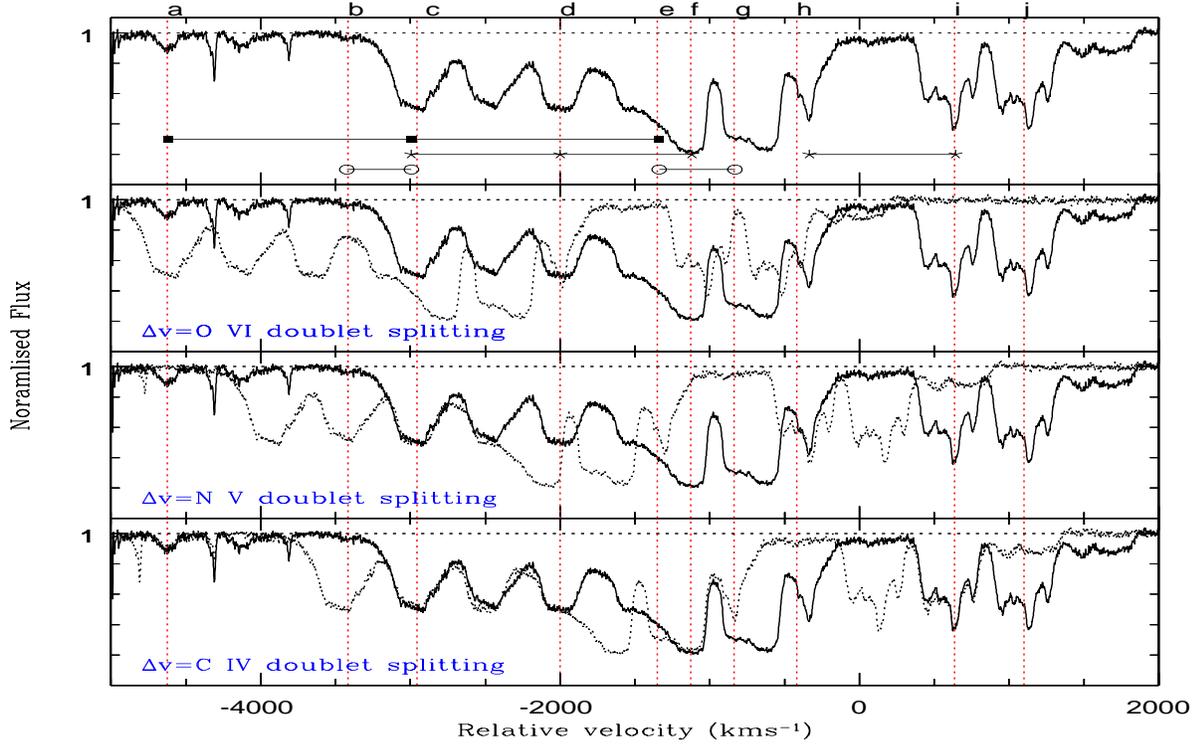,width=16.cm,height=10.cm,clip=,bbllx=25.pt,bblly=702.pt,bburx=580.pt,bbury=158.pt,angle=0.}}} 
\caption[]{C~{\sc iv} absorption profile (top panel and solid line in 
other panels). The same profile, but shifted by different velocity 
separations, is overplotted as a dotted line in the other panels 
(respectively O~{\sc vi}, N~{\sc v} and C~{\sc iv} from top to  bottom). 
Different velocity components that are 
separated by \civ (circle), \nv (star) and \ovi (filled square) doublet 
splittings are marked in the top most panel. 
The vertical dotted lines mark the position of various distinct
component identified in the flow.
}
\label{fig2}
\end{figure*}
\section{Covering factor}
The UVES spectrum of Q~1511+091 over the  C~{\sc iv} and Si~{\sc iv} 
emission line wavelength ranges is shown in the left-hand side panels of 
Fig.~\ref{fig1} together with the fit to the effective continuum
(i.e., quasar continuum plus broad emission lines) we used to normalize 
the spectrum. Partial coverage of the absorbing
gas implies some residual at the bottom of absorption lines 
even if they are heavily saturated. Partial coverage
is apparent when the two resolved lines of a doublet, with different
oscillator strengths, have the same apparent residuals without going to zero. 
This is obviously the case for most of the components in the C~{\sc iv} flow 
toward Q~1511+091. The two absorption lines of the C~{\sc iv} doublet 
corresponding to components {\bf a}, {\bf c}, {\bf d} and {\bf f} have exactly 
the same apparent depths over their profiles.  It can be seen on 
right panels of  Fig.~\ref{fig1} that when the spectrum is shifted
by the velocity separation of the two lines of the C~{\sc iv} doublet, the
profiles coincide nearly perfectly for most of the wavelength range.
Using the method described in Srianand \& Shankaranarayanan (1999)
we estimated the covering factor for different species in individual
components. The results are given in Table.~\ref{tab1}.
\par\noindent
It is possible to say that different part of the flow cover
the same background region. Indeed, it can be noted that when 
\civb of one component, say component 1, overlaps with \civa of 
another component, say component 2, 
the resultant optical depth, although smaller than 0.5 is not the sum 
of the optical depths of the two absorption lines. 
This is particularly striking for \civa of component {\bf e} and \civb 
of component {\bf d} or for \civa of component {\bf g} and \civb of 
component {\bf e}. 
This can be understood if one  of the saturated components, say 
component 1, is shielding the other component, say component 2
in such a way that the flux supposed to reach component 2 in
\civa is absorbed completely by component 1 in C~{\sc iv}$\lambda$1550.
This explains why, although blending is apparent, the \civa profile matches so 
closely the same profile shifted by the velocity splitting between the 
two \civ lines (see Fig.~\ref{fig2}). This shows that at least part 
of the flow is collimated and covers the same part of the emitting region.

\section {Signatures of line-driven acceleration}

\begin{table}
\caption {Velocity splitting among distinct components}
\begin{tabular}{cccc}
\hline
{Doublet} & \multicolumn {1}{c}{splitting} & {components}& separation\\
          & {(\kms)}    &             &  (\kms)   \\
\hline
\civ      &  499        & {\bf b - c} & 495\\
          &             & {\bf e - g} & 499\\
\nv       &  962        & {\bf c - d} & 969\\
          &             & {\bf d - f} & 930\\
          &             & {\bf e - h} & 924\\
\ovi      & 1650        & {\bf a - c} &1660\\
          &             & {\bf c - e} &1625\\
\hline
\end{tabular}
\label{tab2}
\end{table}

The most striking observation in this flow is that the velocity
differences between distinct components coincide with various 
doublet velocity splittings (see Fig~\ref{fig2} and Table.~\ref{tab2}). 
The most apparent 
coincidences are seen for the \nv doublet splitting. 
Both the \civ lines of components {\bf c} and {\bf d} match very closely.
Not only the velocity difference between the centroids of the two components
is very close to 
the \nv doublet splitting but also the
residual intensities and the widths of the components 
($FWHM$~$\sim$~300~km~s$^{-1}$) are nearly identical. 
Very good matchings are also seen  for
one of the lines of {\bf e} and {\bf h} 
and {\bf d} and {\bf f}. 
In addition to this we notice that 
components {\bf h} (defined by weak albeit broad 
\civb and \ovi absorptions)  and {\bf j} are separated by the \nv doublet splitting
(see Fig.~\ref{fig2}).
For the \ovi splitting, 
coincidences are seen 
for components {\bf a} and {\bf c}, 
{\bf c} and 
{\bf e} (with similar residuals). 
This is complemented by coincidences with \civ splitting 
 between {\bf b} and {\bf c}
but mostly between {\bf e} and {\bf g} 
and between component {\bf i} and diffuse absorption component {\bf j}. 
\par\noindent
In a spectral range where crowding of absorption lines is so important,
chance coincidences are possible. In order to consider the probability
that such coincidences occur by chance, we randomly populated the observed 
velocity range with 10 components. Using 10$^6$ realizations we find a probability of
6$\times10^{-4}$ for the occurrence of, at least two coincidences, each 
with the velocity splittings of the \ovi, \nv or \civ doublets with a  
matching uncertainty of $\pm$20 \kms. 
If we introduce the additional condition that
at least two coincidences occur between three components in \ovi
splitting (like the one seen among {\bf a}-{\bf c} and {\bf c}-{\bf e})
the probability becomes $<10^{-6}$ for a wavelength
matching uncertainty of $\pm$20 \kms. 
\par\noindent
This low probability for mere velocity coincidence together with tight 
matching of some of the profiles and consistent covering factors clearly 
confirms the line-locking situation in this system.
It must be noted that, although the corresponding wavelength range is 
crowded by intervening lines, additional coincidences are apparent 
between the corresponding \ovi absorptions.
In addition, the velocity separation between components {\bf a} and the 
red component of {\bf i} ($\Delta_v= 5830$~km~s$^{-1}$) is 
very close to the \nv to \lya velocity splitting. 
Therefore, all components are 
connected by a complex set of coincidences.
Under the assumption that line-locking is structuring the flow,
it is expected that components with larger covering factors
be located closer to the source. This would imply in that case
that the flow is an accelerating wind.

\section{Discussion}
\begin{figure}
\centerline{\vbox{
\psfig{figure=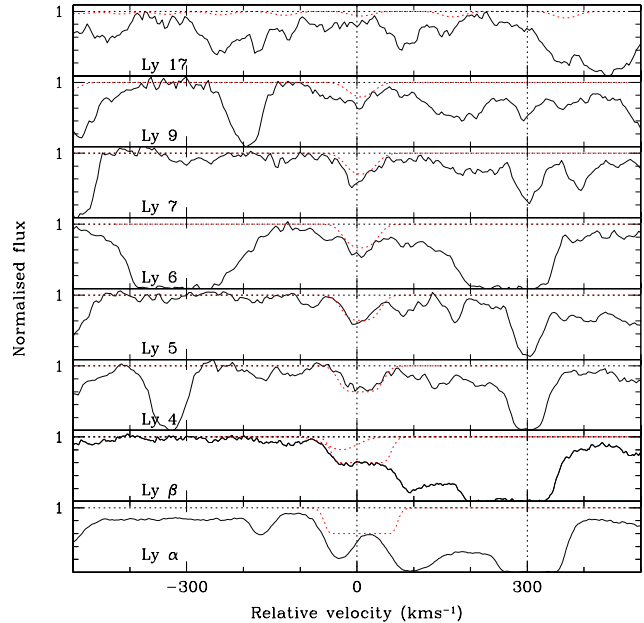,height=9.cm,width=9.cm,angle=0.}
}}
\caption[]{
Velocity plot of Lyman series absorption lines produced by 
component {\bf e} at $z_{\rm abs}$~= 2.8606 toward Q~1511+091. 
The best fitted profile (dotted line) is
obtained considering the gas covers only 40\% of the continuum 
emission (from the accretion disk). Note that a weak additional 
intervening Lyman $\alpha$ line is blended with the Lyman $\beta$ line. 
The effect of partial coverage can be appreciated by comparing the 
profiles of this component with that of an intergalactic absorption 
with similar amount of H~{\sc i} seen at 300 \kms.
}
\label{fig3}
\end{figure}

\par\noindent

\begin{figure}
\centerline{\vbox{
\psfig{figure=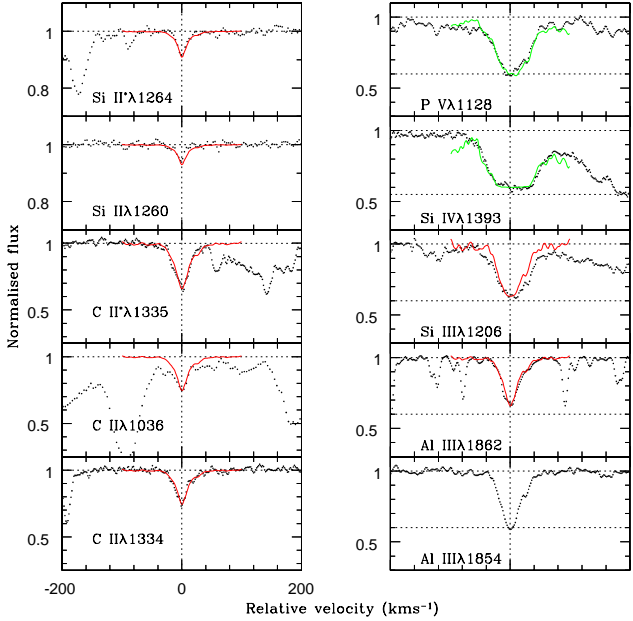,height=9.cm,width=9.cm,angle=0.}
}}
\caption[]{
Fits of the absorption profiles in component 
{\bf e} at $z_{\rm abs}$~= 2.8606 toward Q~1511+091. The observed profile of 
Al~{\sc iii}$\lambda$1854 is used as a template to fit the singly 
ionized species. The profile are fitted assuming a covering factor
of 0.4 and by varying the central optical depth preserving the 
profile function given by the template.
P~{\sc v} and Si~{\sc iv} profiles are fitted using the Si~{\sc iii} 
profile as the reference template. The difference between the two profiles 
is mainly due to an extra component in the red wing that is seen in 
Si~{\sc iii} and other higher ionization lines. 
}
\label{fig4}
\end{figure}

The physical conditions in part of the flow can be investigated 
from the very peculiar component {\bf e} at \zabs~=~2.8606 
($-$1350~km~s$^{-1}$ in Fig.~\ref{fig2}). Absorption lines from
H~{\sc i} Lyman series lines as well as excited levels 
of singly ionized species \cii and \siii are present (see Fig.~\ref{fig3} and
from C~{\sc ii} to P~{\sc v}. 
However, none of the Lyman series  absorption lines
are dark. Lyman-$\alpha$ is not even seen as a distinct component when it should be conspicuous. 
Similar situation was observed by Telfer et al. (1998) toward
QSO SBS 1542+531. From the flat bottom of the Lyman-$\beta$ line, it 
is apparent that the cloud does not cover the background source completely.
Higher Lyman series lines are well fitted  with a covering factor
of the order of 0.4. 
It must be noted that the emission lines for the corresponding transitions
are negligible compared to the continuum. In addition,
such a low covering factor can not be accommodated by scattered light (Ogle et al. 1999) . 
Therefore, it can be concluded that {\sl the cloud covers only 
about 40\% of the central source of continuum} whose radius is of the 
order of 10$^{-3}$~pc.
All neutral hydrogen lines can be fitted consistently (see
Fig.~\ref{fig3}) with $f_{\rm c} = 0.4$ and log~$N$(H~{\sc i})~$\sim$~16.30.
A consistent model for all other absorption lines is found with this covering factor
(see Fig.~\ref{fig4}) and using the Al~{\sc iii} and Si~{\sc iii} profiles as 
templates for, respectively, the singly-ionized and high-ionization species.
The difference between the two profiles 
is mainly due to an extra component in the red wing that is seen in 
Si~{\sc iii} and other higher ionization lines. 
Note that although the centroids of the lines are the same, the width of
the lines increases with ionization. 
\par\noindent
It is striking to note that with this
covering factor, the ratios C~{\sc ii}$^*$/C~{\sc ii} and 
Si~{\sc ii}$^*$/Si~{\sc ii} are found to be both of the order of two which
is the value expected in case of thermal equilibrium. Given the ionization
state of the gas
(see below), the temperature is most probably larger than ~10$^4$~K, and 
from the thermal equilibrium between populations of Si~{\sc ii} levels we 
can derive the electronic density 
$n_e\ge 1.3\times$10$^3$~$T^{0.5}_4$~cm$^{-3}$
(T$_4$ is the temperature in the units of 10$^4$ K). 
For a Mathews \& Ferland (1987) QSO ionizing spectrum, and using
the constraints from observed ion ratios, 
the ratio of the ionizing photon density to the hydrogen particle 
density (the so-called ionization parameter, U) can be shown to be in the
range $-$2.0~$<$~log~$U$~$<$~$-$0.6. 
The metallicities are most probably larger than 
solar. This is somewhat consistent with the typical values obtained for
the Broad Line Region, $-$1.0~$<$~log~$U$~$<$~$-$0.0
and $10^9\le n_{\rm e}$~$\le 10^{11}$~cm$^{-3}$, although the density is
probably smaller in the cloud (see below).
Assuming a typical luminosity density at the Lyman limit 
of $10^{31}~{\rm erg~s^{-1}~Hz^{-1}}$, and a spectral index of 
$-1.6$ the maximum distance of the absorbing gas from the source
is $~15.75\sqrt{ 1/ n U}~{\rm kpc}$. The lower limit on $n_e$
derived above suggests
the gas has to be at a distance less than a few kpc from the 
ionizing source. 
\par\noindent
From the absence of absorption due to C~{\sc iii}$^{*}\lambda1176$
originating from an excited metastable level,
we derive an upper limit on the density of 8$\times10^9$ cm$^{-3}$.
This gives a minimum distance of a few 10$^{-1}$ pc. Thus
the observations are consistent with the absorbing gas being inside or
close to the BLR.
\begin{figure*}
\centerline{\vbox{
\psfig{figure=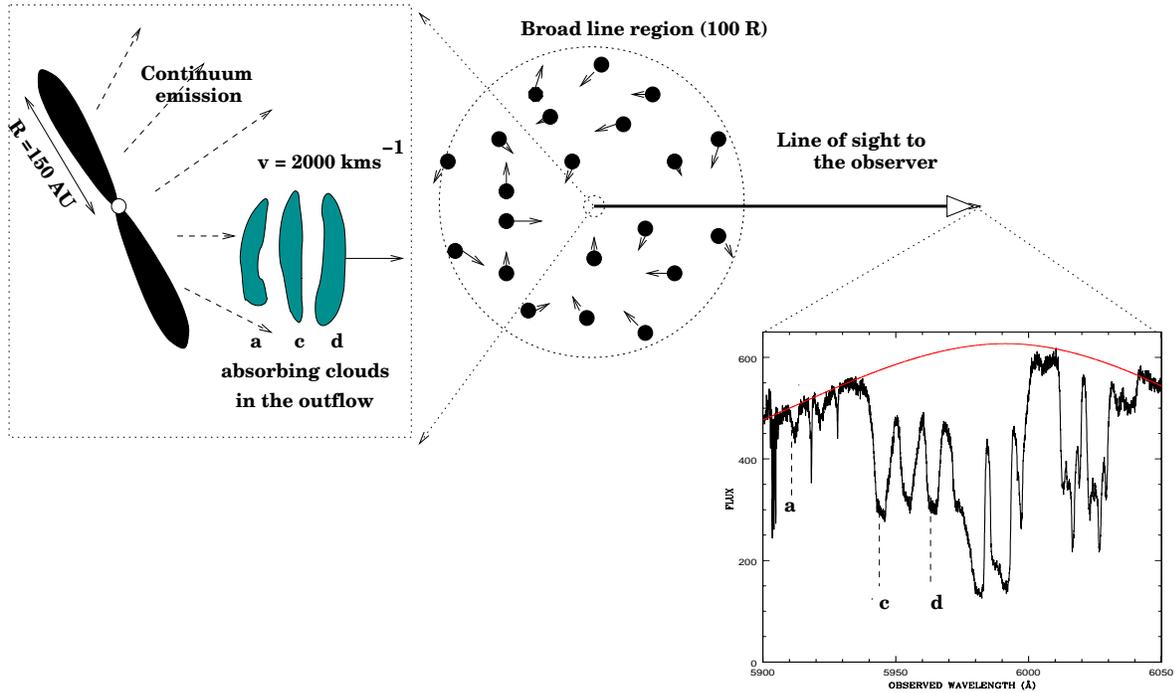,width=16.cm,height=10.cm,clip=,bbllx=90.pt,bblly=81.pt,bburx=500.pt,bbury=750.pt,angle=270.}}}  
\caption[]{
Qualitative sketch of the relative position
of the absorbing gas compared to the flux emitting region.
The fact that some of the components of the outflow from Q~1511+091
cover only part of the source of continuum 
suggests that the absorbing gas is close to the continuum emission region.
Typically the continuum is believed to originate from a region of
size $R$~$\sim$~150~AU. Reverberation mapping studies of AGNs indicate 
that the broad emission lines are
produced by a region of radius at least 100 times that of the continuum
emitting region. Thus it is most likely that at least part of
the observed flow
originate close to the inner boundary of the broad emission
line region.
}
\label{fig5}
\end{figure*}
For a spherical absorbing cloud the particle density is
$\sim N$(H)/$\sqrt{f_{\rm c}} R_{\rm c}$, where $R_{\rm c}$ and
$N$(H) are the radius of the continuum emitting region and the total 
hydrogen column density of the absorbing gas 
(in the range, $10^{19}\le$ $N$(H)$\le10^{20}$ cm$^{-2}$,
from the models) respectively.
For the lower limit on density we get, $R_{\rm c}\leq0.04~T_4^{-0.5}$pc.
\par\noindent
The largest blue-shift observed in the C~{\sc iv} profile is much 
smaller than the typical terminal velocities of BAL flows. 
This probably means that the gas could be further accelerated.
Indeed, for physical conditions similar to those observed in 
different components of the Q~1511+091 outflow, the line-radiation force is
much larger than gravity even in the close proximity of the
black-hole (de Kool \& Begelman 1995) .
Most of the hydrodynamical  radiatively driven wind
models predict the gas to reach terminal velocity at a distance
of a few pc from the ionizing source (Arav et al. 1994).
Lack of acceleration in components with velocity much lower
than the terminal velocity could be accounted for if either our
line of sight probes a small portion of a non-radial flow
(Arav 1996) or the extra drag forces are at play in reducing 
the  effective acceleration of the gas. 
In the case of Q~1511+091, the absorption lines are seen at
redshift very close to or slightly higher than the systemic redshift.
In addition, nearly perfect velocity matching with doublet splittings
suggests that the direction of the flow is very close
to our line of sight.
\par\noindent
Disk winds have been suggested as the natural origin for ejected material
from quasars giving rise to BAL systems (de Kool \& Begelman 1995;
Murray et al. 1995; Proga et al. 2000). In these models, a cylindrically symmetric 
disk wind is ejected away from the surface of the disk and is radiatively 
accelerated radially once exposed to the strong continuum source.
Line-driven instabilities could lead to  the formation of distinct velocity 
components seen in the flow (Feldmeier et al. 1997).
Our observations illustrated in Fig.~\ref{fig5} 
support the basic mechanism at work in these models. 
It would be very interesting to constrain the acceleration over a large
time-scale
for our understanding of the nature of drag forces and in
general the nature of the flow.

%
\section*{acknowledgments}
We thank the referee, Dr. N. Arav, for useful comments.
We gratefully acknowledge support from the Indo-French Centre for 
the Promotion of Advanced Research (Centre Franco-Indien pour la Promotion
de la Recherche Avanc\'ee) under contract No. 1710-1. This work 
was supported in part by the European Communities RTN network
"The Physics of the Intergalactic Medium". RS thanks the Institute of
Astronomy in Cambridge and the Institut d'Astrophysique de Paris
and PPJ thanks IUCAA for hospitality during the time part of this work 
was completed.
%
%


\begin{thebibliography}{}
\bibitem{}
Arav N., 1996, ApJ, 465, 617
\bibitem{}
Arav N. \& Begelman M., 1994, ApJ, 434, 479
\bibitem{}
Arav N., Li Z., Begelman M., 1994, ApJ, 432, 62
\bibitem{}
Barlow T. A. \& Sargent W. L. W., 1997, AJ, 113, 136
\bibitem{}
Braun E. \& Milgrom M., 1989, ApJ, 342, 100
\bibitem{}
Castor J. I., Abbott D. C., Klein R. I., 1975, ApJ, 195, 157.
\bibitem{}
de Kool M. \& Begelman M. C, 1995, ApJ, 455, 448
\bibitem{}
D'Odorico S., et al., 2000, SPIE, 4005,1 
\bibitem{}
Feldmeier A., Norman C., Pauldrach A., Owocki S., Puls J.,
Kaper, L., 1997, in Mass ejection from Active Galactic Nuclei,
ASP Conference Series, Vol, 128, p258.
\bibitem{}
Foltz C. B., Weymann R. J., Morris S. L., Turnshek D. A., 1987, 
ApJ, 317, 450
\bibitem{}
Hamann F., 1997, ApJS, 109, 279
\bibitem{}
Hutchings J. B., et al., 2001, ApJ, 559, 173
\bibitem{}
Korista T. K., Voit G. M., Morris S. L., Weymann
R. J., 1993, ApJS,  88, 357
\bibitem{}
Mathews W. G. \& Ferland G. J., 1987, ApJ, 323,456
\bibitem{}
Murray N., Chiang J., Grossman A., Voit M., 1995, ApJ. 441, 498
\bibitem{}
Ogle P. M., Cohen M. H., Mill J. S., Tran H. D., Goodrich
R. W., Martel A. R., 1999, ApJS, 125, 1
\bibitem{}
Petitjean P., Rauch M., Carswell R. F., 1994, A\&A, 291,29
\bibitem{}
Proga D., Stone J. M., Kallman T. R., 2000, ApJ, 543, 686
\bibitem{}
Sargent, W. L. W., Steidel, C. C., \& Boksenberg, A. 1988, ApJS, 68, 539
\bibitem{}
Scargle J. D., 1973, ApJ, 179, 705
\bibitem{}
Srianand R., 2000, ApJ, 528, 617
\bibitem{}
Srianand R. \& Petitjean P., 2001, A\&A, 357, 414
\bibitem{}
Srianand R. \& Shankaranarayanan, 1999, ApJ, 518, 672
\bibitem{}
Telfer R. C., Kriss G. A., Zheng W., Davidsen A. F., Green R. F., 1998, ApJ, 509, 132 
\bibitem{}
Vilkoviskiji E. Y. \& Irwin M. J., 2001, MNRAS, 321, 4
\bibitem{}
Weymann R. J., Morris S. L., Foltz C. B., Hewett P. C., 1991, ApJ, 373, 25
\end{thebibliography}
\end{document}